\documentclass{eas}
\usepackage{graphicx}
\begin{document}

\title{A common behavior in the late X-ray afterglow of energetic GRB-SN systems} 
\author{L. Izzo}
\address{Sapienza University of Rome, P.le Aldo Moro 5, I-00185 Rome, Italy, \email{luca.izzo@gmail.com}}
\secondaddress{ICRANet, Piazza della Repubblica 10, I-65122 Pescara, Italy}
\author{G. B. Pisani}
\address{Erasmus Mundus Joint Doctorate IRAP PhD. Student. Universit«e de Nice Sophia Antipolis, Nice, CEDEX 2, Grand Chateau Parc Valrose}
\author{M. Muccino}\sameaddress{1}
\author{J. A. Rueda}\sameaddress{1,2}
\author{Y. Wang}\sameaddress{1}
\author{C. L. Bianco}\sameaddress{1,2}
\author{A. V. Penacchioni}\sameaddress{3}
\author{R. Ruffini}\sameaddress{1,2}
\begin{abstract}
The possibility to divide GRBs in different subclasses allow to understand better the physics underlying their emission mechanisms and progenitors. The induced gravitational collapse scenario proposes a binary progenitor to explain the time-sequence in GRBs-SNe. 
We show the existence of a common behavior of the late decay of the X-ray afterglow emission of this subclass of GRBs, pointing to a common physical mechanism of their late emission, consistent with the IGC picture. 
\end{abstract}
\maketitle
It has been proposed that the temporal coincidence of a Gamma Ray Burst (GRB) and a supernova (SN) Ib/c emission (GRB-SN) can be explained by the concept of induced gravitational collapse (IGC) (Ruffini et al. 2001,2007). This concept has been extended recently \cite{3} and can be summarized as follows. In the IGC scenario the GRB-SN progenitor is a binary system composed by an evolved massive star and a Neutron Star(NS). The evolved star undergoes a SN explosion leading to a subrelativistic expansion of the SN core progenitor outer layers, while its high density core contracts, as it was shown in \cite{5}. Since the SNe associated to GRBs are of type Ib/c, the SN core progenitor is very likely an evolved Wolf-Rayet, or a Carbon-Oxygen (CO) core. Part of the expelled material is accreted at a high rate by the NS companion, fastly increasing the NS mass. The NS can reach, in a few seconds, the critical mass and consequently gravitationally collapses to a Black Hole (BH). This gravitational collapse process leads to the emission of a canonical GRB. The SN emission may again be observed as an optical bump in the late afterglow emission. 

These two distinct emissions, the early SN expansion that leads to the accretion process (Episode 1) and the actual GRB emission (Episode 2), have been called double emission episodes and were clearly identified in the gamma-ray energy range in GRB 090618 \cite{6} and GRB 101023 \cite{8}. Recent observations of the late ($t=10^8-10^9$ s) emission of low energetic GRBs-SNe, or X-ray Flashes (XRFs), show a distinct emission in the X-ray regime consistent with temperatures $T \sim 10^7$--$10^8$ K. Similar features have been also observed in the two Type Ic SNe (SN 2002ap and SN 1994I) that are not associated to GRBs. It was shown \cite{7} that this late decay emission in the X-rays of GRB 980425/sn1998bw, GRB030329/sn2003dh and GRB03123/SN2003lw might be explained as the luminosity coming from the cooling of the neo-NS from the remnant of the SN. In more energetic GRBs-SNe additional mechanisms, related either to the BH, formed from the IGC of the NS companion, or to the neo-NS left by the SN, could be at work. We will present elsewhere these other mechanisms in connection with the IGC scenario. 

We turn now to the analysis of the late X-ray afterglow of our sample of GRBs-SNe. It consists of 6 GRBs, with redshift, for which a SN event was observed after about 10 days from the GRB trigger, or a SN was not observed but there is a double emission episode in the prompt emission of the GRB. We have also considered two additional GRBs, 101023 and 110709B, for which there is no a redshift observation, but their prompt emission shows evidence of a double emission episode. 
For these two GRBs there is an estimated redshift of 0.9 and 0.75 respectively \cite{8},\cite{16}.  
The sample include GRB 090618 and GRB 060729, at redshift of 0.54, for which a photometric bump associated to a SN event was detected and reported in literature \cite{9}. GRB 091127 is associated with sn2009nz at a redshift of $z=0.49$. We include also GRB 111228 for which a transient, associated with a SN event, was detected in the differential photometry of the optical emission in two epochs, 34 and 76 days after the GRB trigger \cite{10}. We made similar considerations for GRB 080319B, where the optical emission in the $i'$-band shows a decay unusual for a GRB afterglow and more related to an underlying SN emission \cite{11}. We include also GRB 061007, for which no SN event was detected, due to the lack of late optical observations, but for which a clear thermal first emission episode was reported in literature \cite{12}. These GRBs have an isotropic energy larger than $10^{52}$ erg, which is at least two orders of magnitude larger than the energy emitted in the well-known cases of GRB 980425 and GRB 060218, which have a relative low-redshift and are not taken into account in this analysis.

We have compared the late X-ray emission of all GRBs in the sample. We developed a code which allows to transform this late GRB emission in a luminosity (0.3-10 keV) light curve computed in the rest-frame. In order to extrapolate any spectrum in a common rest-frame, we have corrected any GRB for a specific correction factor, proportional to the respective redshift. After the correction for the distance and for the time, we have included all the X-ray afterglow light curves in a common picture, see left panel in Fig.1. It is evident a common behavior at late times, $t > 10^4$ s. A fit with a simple power-law function, $L_{t_0} \propto t^{\gamma}$ of the emission after $t_0=2\times 10^4$ s at $z=1$, provides the values of the luminosity at the initial time $t_0$ and of the time decay index. We noted a clustering of the distribution of the luminosities for any GRB in the sample around a best-fit line, similar to other results in literature (Penacchioni et al. 2012, 2013). We then estimated the redshift for GRB 101023,  $0.6 < z < 1.2$ with best value $z = 1.0$ (see right panel in Fig. 1), and GRB 110709B,  $0.4 < z < 0.6$ with best value $z = 0.5$, by varying their redshift and computing the corresponding $L_{t_0}$ and $\gamma$.  

\begin{figure}
\includegraphics[scale=0.183]{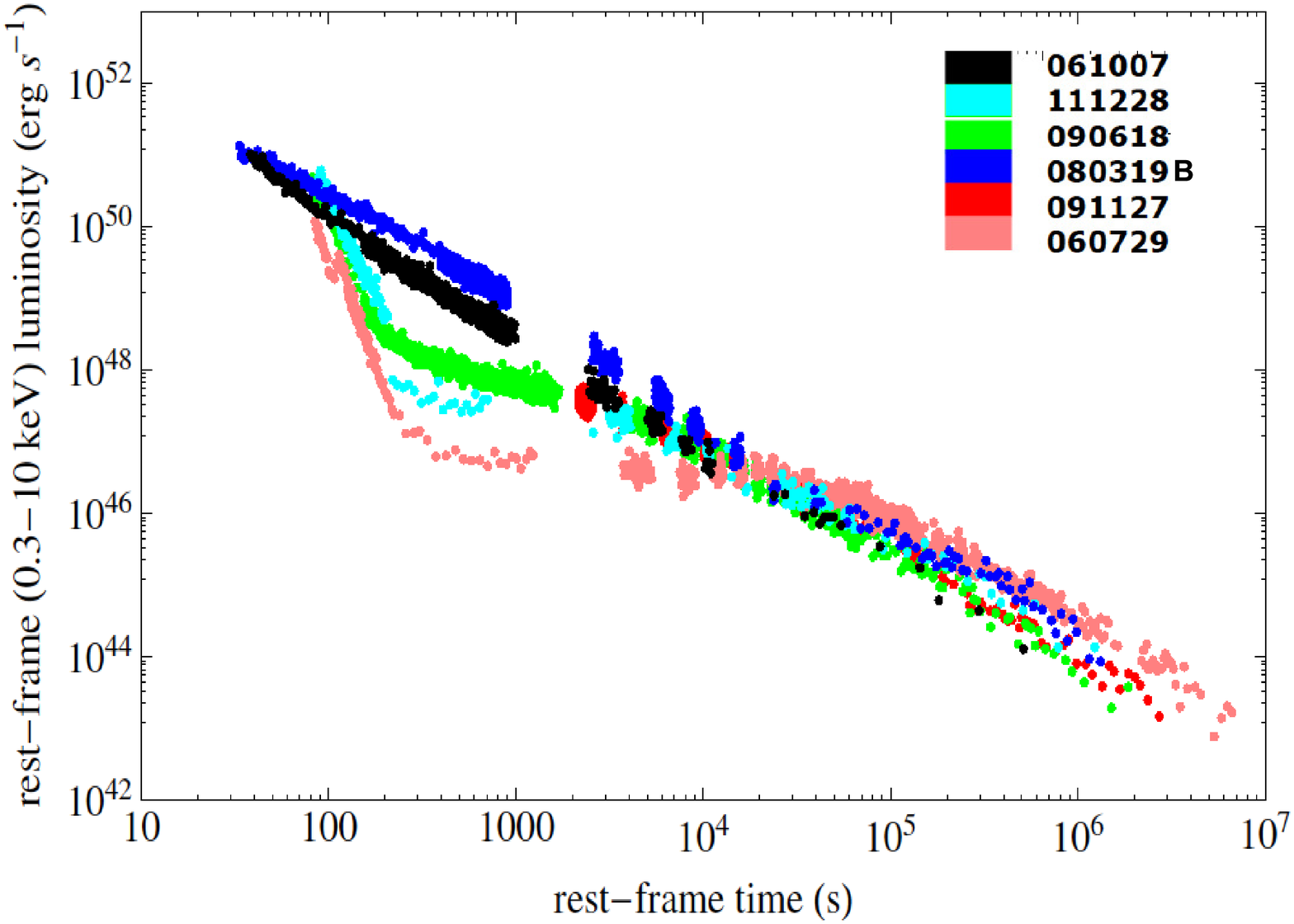}
\includegraphics[scale=0.32]{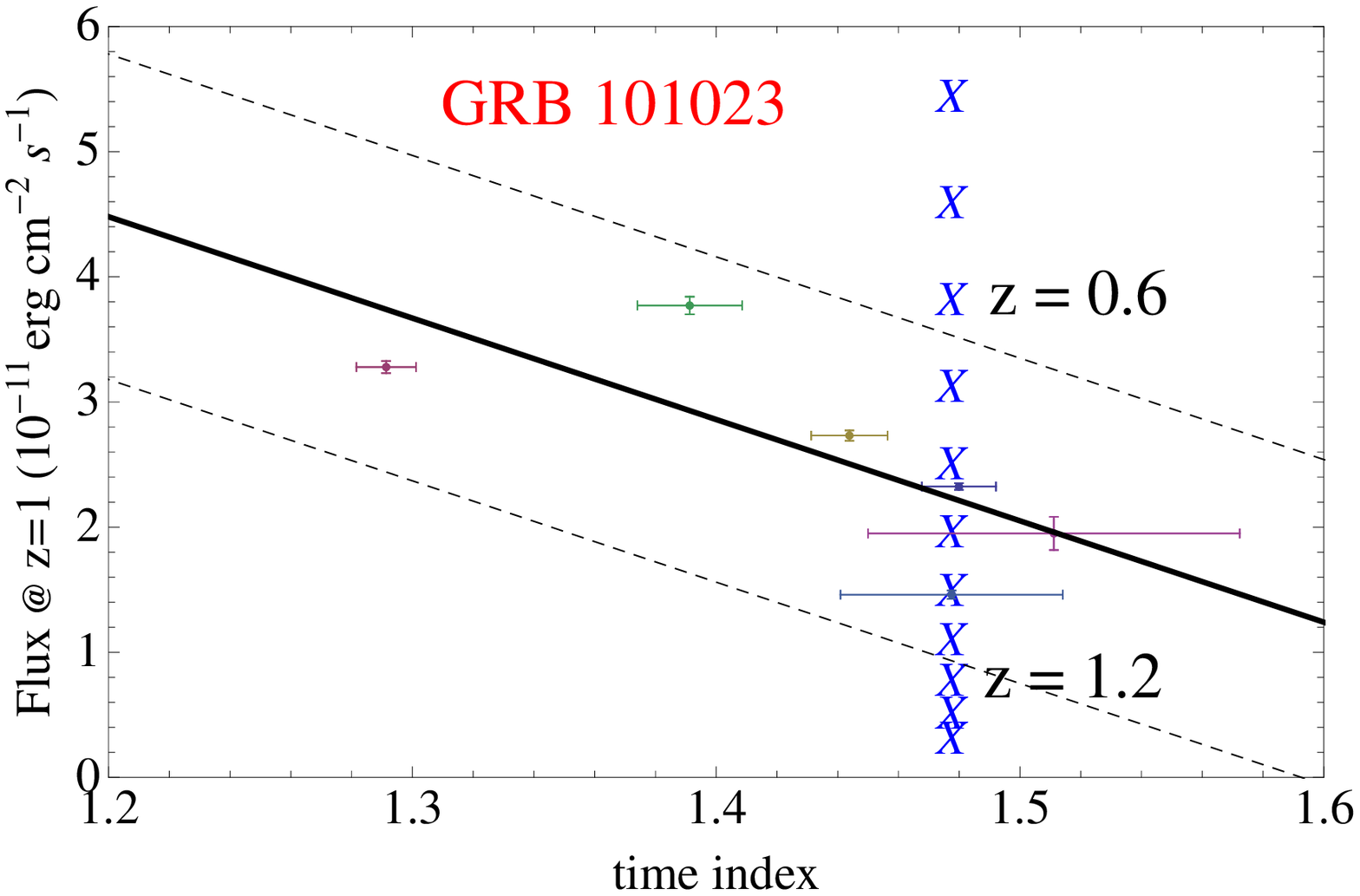}
\caption{The X-ray luminosity of the six GRBs with measured redshift in the $0.3\,$--$\,10$ keV rest-frame energy range (left panel). The estimates of the redshift for GRB 101023. The solid black line is the best fit of the correlation $F_{obs,z=1}$-$\gamma$ where $F_{obs,z=1}$ is the observed flux at z=1 and at t=20000 s in the rescaled light curve, $\gamma$ is the power-law index decay of the late X-ray afterglow light curve. The dashed lines correspond to a deviation of 2$\sigma_{ext}$ from the best fit line, where $\sigma_{ext}$ is the extra scatter error computed using the method explained in \cite{13}. The blue crosses represent the values of ($F_{obs,z=1}$,$\gamma$) for GRB 101023 computed for different values of the redshift (right panel).} 
\end{figure}


\end{document}